\documentclass{aastex61}
\usepackage[utf8]{inputenc}
\usepackage{amsmath,amsfonts,amssymb}
\usepackage{subfigure}
\usepackage{epstopdf}
\usepackage{graphicx,enumerate}
\usepackage{hyperref}
\usepackage{enumitem}
\usepackage{natbib}
\usepackage{xcolor}

\shorttitle{OVRO-LWA: Search for GRB prompt radio emission}
\shortauthors{Anderson et al.}

\begin{document}
\title{A simultaneous search for prompt radio emission associated with the short GRB 170112A using the all-sky imaging capability of the OVRO-LWA}

\author{Marin M. Anderson}
\affiliation{California Institute of Technology, 1200 E California Blvd MC 249-17, Pasadena, CA 91125, USA}

\author{Gregg Hallinan}
\affiliation{California Institute of Technology, 1200 E California Blvd MC 249-17, Pasadena, CA 91125, USA}

\author{Michael W. Eastwood}
\affiliation{California Institute of Technology, 1200 E California Blvd MC 249-17, Pasadena, CA 91125, USA}

\author{Ryan M. Monroe}
\affiliation{California Institute of Technology, 1200 E California Blvd MC 249-17, Pasadena, CA 91125, USA}

\author{Harish K. Vedantham}
\affiliation{California Institute of Technology, 1200 E California Blvd MC 249-17, Pasadena, CA 91125, USA}

\author{Stephen Bourke}
\affiliation{California Institute of Technology, 1200 E California Blvd MC 249-17, Pasadena, CA 91125, USA}
\affiliation{Department of Space, Earth and Environment, Chalmers University of Technology, Onsala Space Observatory, S-439 92 Onsala, Sweden}

\author{Lincoln J. Greenhill}
\affiliation{Harvard-Smithsonian Center for Astrophysics, 60 Garden Street, Cambridge MA 02138 USA}

\author{Jonathon Kocz}
\affiliation{California Institute of Technology, 1200 E California Blvd MC 249-17, Pasadena, CA 91125, USA}

\author{T. Joseph W. Lazio}
\affiliation{Jet Propulsion Laboratory, California Institute of Technology, 4800 Oak Grove Dr, Pasadena, CA 91109, USA}

\author{Danny C. Price}
\affiliation{Harvard-Smithsonian Center for Astrophysics, 60 Garden Street, Cambridge MA 02138 USA}
\affiliation{Centre for Astrophysics \& Supercomputing, Swinburne University of Technology, PO Box 218, Hawthorn, VIC 3122, Australia}

\author{Frank K. Schinzel}
\affiliation{National Radio Astronomy Observatory, P.O. Box O, Socorro, NM 87801 USA}
\affiliation{Department of Physics and Astronomy, University of New Mexico, Albuquerque, NM 87131 USA}

\author{Yuankun Wang}
\affiliation{California Institute of Technology, 1200 E California Blvd MC 249-17, Pasadena, CA 91125, USA}

\author{David P. Woody}
\affiliation{California Institute of Technology, Owens Valley Radio Observatory, Big Pine, CA 93513, USA}

\correspondingauthor{Marin M. Anderson}
\email{mmanders@astro.caltech.edu}

\begin{abstract}
We have conducted the most sensitive low frequency (below 100~MHz) search to date for prompt, low-frequency radio emission associated with short-duration gamma-ray bursts (GRBs), using the Owens Valley Radio Observatory Long Wavelength Array (OVRO-LWA). The OVRO-LWA's nearly full-hemisphere field-of-view ($\sim20$,$000$ square degrees) allows us to search for low-frequency (sub-$100$~MHz) counterparts for a large sample of the subset of GRB events for which prompt radio emission has been predicted. Following the detection of short GRB 170112A by \textit{Swift}, we used all-sky OVRO-LWA images spanning one hour prior to and two hours following the GRB event to search for a transient source coincident with the position of GRB 170112A. We detect no transient source, with our most constraining $1\sigma$ flux density limit of $650~\text{mJy}$ for frequencies spanning $27~\text{MHz}-84~\text{MHz}$. We place constraints on a number of models predicting prompt, low-frequency radio emission accompanying short GRBs and their potential binary neutron star merger progenitors, and place an upper limit of $L_\text{radio}/L_\gamma \lesssim 7\times10^{-16}$ on the fraction of energy released in the prompt radio emission. These observations serve as a pilot effort for a program targeting a wider sample of both short and long GRBs with the OVRO-LWA, including bursts with confirmed redshift measurements which are critical to placing the most constraining limits on prompt radio emission models, as well as a program for the follow-up of gravitational wave compact binary coalescence events detected by advanced LIGO and Virgo.
\end{abstract}

\section{Introduction}\label{introduction}
The detection of the first gamma-ray bursts (GRBs) in 1967 heralded a race to better characterize, classify, and identify the nature of the progenitors of these seconds-long bursts of MeV gamma-rays, which appeared to be isotropically distributed across the sky and thus likely of cosmic origin~\citep{Klebesadel+1973}. Systematic detections of GRBs over the following decades revealed two distinct classes of events: the spectrally hard, short (typical duration $<2$ seconds) GRBs and the spectrally soft, long (typical duration $>2$ seconds) GRBs~\citep{Kouveliotou+1993}. The rapid follow-up capabilities of the BATSE instrument and BeppoSAX led to the detection of X-ray, optical, and, later, radio afterglows that provided critical information regarding distance, host galaxy association, isotropic energy estimates, source size evolution, and insight into GRB progenitors~\citep{Costa+1997,Frail+1997,vanParadijs+1997}. Routine follow-up and afterglow detection revealed that the distinct phenomenology of short and long GRBs also reflects distinct progenitor systems. Long GRB hosts are exclusively star forming galaxies~\citep{Savaglio+2009}, and the location of long GRBs within their host galaxies correlates strongly with ultraviolet light, implying that long GRBs trace regions of active massive star formation~\citep{Fruchter+2006}. This, combined with the association of long GRBs with Type Ic core-collapse supernovae (SNe), points to massive stars as the progenitors of long GRBs~\citep{Woosley+Bloom2006}.

The progenitors of short GRBs, however, have remained more elusive. While significant evidence exists for the association of short GRBs with compact object mergers consisting of neutron star binaries (NS-NS) or neutron star black hole binaries~\citep[NS-BH;][]{Narayan+1992}, the association is not as definitive as that of long GRBs with core-collapse supernovae~\citep[see, e.g.,][]{Lyutikov2009}. However, the compact object merger scenario is the favorable progenitor model for short GRBs for the following reasons: (1) short GRBs are found in both early and late-type galaxies, consistent with the formation of progenitor binary systems following a delay-time distribution and therefore being found in both young and old stellar populations; (2) localization within host galaxy provided by detections of short GRB afterglows indicates a population distribution with significantly larger host galaxy offset relative to the long GRB / core-collapse SN population, as expected for a compact object binary progenitor born with a natal kick~\citep{Bloom+1999, Belczynski+2006}; and (3) unlike long GRBs, short GRBs have no established association with supernovae~\citep{Berger+2005,Bloom+2006,Soderberg+2006,Berger2009}, although they are associated with kilonovae / macronovae believed to be powered by r-process nucleosynthesis in the expanding post-merger ejecta~\citep{Li+Paczynski1998,Tanvir+2013,Yang+2015}.

A number of models predict a highly speculative but potentially very valuable counterpart to GRBs and NS-NS(BH) mergers in the form of a short, bright, coherent pulse of low frequency radio emission (Table~\ref{tab:models}). The models predicting this coherent radio emission span all stages of the compact object merger process, from (1) the final in-spiral of the binary neutron stars; to (2) a short-lived, post-merger supramassive neutron star; to (3) the post-collapse stage during which the gamma-ray emission is produced. \citealt{Hansen+Lyutikov2001} consider the magnetospheric interaction of a NS-magnetar binary system and the generation of a coherent radio burst in the surrounding plasma environment during the pre-merger (1) phase~\citep[see also][]{Lyutikov2013}. \citealt{Pshirkov+Postnov2010} consider a low-frequency radio burst generated in the relativistic plasma outflow from the highly magnetized, rapidly rotating magnetar which is predicted to form in the brief stage (2) between the merger and final collapse. \citealt{Usov+Katz2000} predict a low-frequency radio burst that may be produced in the post-merger (3) phase (as well as in long GRBs) through the interaction of a strongly magnetized wind with the circumburst medium. In this model, the coherent low-frequency emission is produced by the time-variable surface current that exists at the wind/ambient plasma-boundary. Other models predicting coherent radio emission post-merger include synchrotron maser emission generated during the GRB fireball phase~\citep{Sagiv+Waxman2002}, and inverse Compton radiation generated in the surrounding magnetized plasma by magnetohydrodynamic modes excited by the gravitational waves produced in the merger~\citep{Moortgat+Kuijpers2005}.

Despite the diversity of models, common to all is the prediction that a GRB is accompanied by a coherent and intrinsically short-duration burst of radio emission that occurs within a window of several seconds to the production of the gamma-ray emission, with a steep negative spectral index that favors observations at lower frequencies. In the case of \citealt{Usov+Katz2000}, the emission is predicted to peak at $\sim$MHz frequencies and fall off rapidly above roughly $30~\text{MHz}$. Many of the models also require the presence of extreme, magnetar-strength magnetic fields (e.g. as high as $10^{15}~\text{G}$), which is atypical given the expected age of these systems at the time of merger~\citep{Goldreich+Reisenegger1992}, but which are justified in the models through magnetic field amplification during the coalescence of the system~\citep[and through numerical simulations, e.g.,][]{Duez+2006}.

\begin{deluxetable}{cCCCCCCC}
	\tabletypesize{\scriptsize}
	\tablecolumns{7}
	\tablecaption{Models for prompt radio counterparts to GRBs\label{tab:models}}
	\tablehead{ 
	(1)					& (2)								& (3)							& (4)					& (5) 				& (6) 			& (7) 					\\
	\colhead{Reference(s)}	& \colhead{Delay relative to GRB}		& \dcolhead{t_{\text{intrinsic,R}}}	& \colhead{Flux density} 	& \colhead{Spectral index}& \colhead{B-field}	& \dcolhead{L_R/L_\gamma}	\\
						&								&							& \colhead{[Jy]}			& 					& \colhead{[G]}		& 						}
	\startdata
	\citealt{Hansen+Lyutikov2001} / \citealt{Lyutikov2013}\tablenotemark{a}	& \propto -t^{-1/4}						& \propto -t^{-1/4}			& 0.07 			& -1 		& 10^{12} 			& 10^{-10}\propto\epsilon_R	\\
	\citealt{Pshirkov+Postnov2010}\tablenotemark{b} 					& \sim10\text{s of ms}~\text{before} 			& 10-100~\text{ms} 			& 9\times10^{4} 	& -2 		& 10^{14} - 10^{16} 	& 10^{-5}\equiv\eta			\\
	\citealt{Usov+Katz2000}\tablenotemark{c} 						& \text{simultaneous} 					& t_{\gamma,\text{prompt}} 	& 350 			& -0.6 	& 10^{15} - 10^{16} 	& 10^{-4}\equiv\delta			\\
	\enddata
	\tablenotetext{a}{Coherent radio emission is generated in the pre-merger phase through the magnetospheric interactions between the two components of the binary system; the resulting magnetized wind produces coherent radio emission with luminosity that scales as $t^{-1/4}$ until it peaks at the time of merger, a few ms prior to the production of the GRB. Flux density prediction in column (4) is calculated at $30~\text{MHz}$ at the peak of emission, and assumes the efficiency with which wind power is converted into coherent radio emission is $\epsilon_R=10^{-5}$. See Equation 13 in~\citealt{Lyutikov2013}.}
	\tablenotetext{b}{Coherent radio emission is generated by the rapid conversion of rotational energy into magnetic energy in the rotationally-supported massive NS, which forms following the merger of the BNS system, but prior to its final collapse and the generation of the GRB. Flux density prediction in column (4) is calculated at $30~\text{MHz}$, and assumes a conversion coefficient $\eta$, defined by the authors as the efficiency with which spin-down energy is converted into radio luminosity, as $\eta=10^{-5}$. See Equation 7 in~\citealt{Pshirkov+Postnov2010}.}
	\tablenotetext{c}{Coherent radio emission occurs simultaneous to gamma-ray emission, and is powered by Langmuir waves generated at the boundary between the highly magnetized outflow and the surrounding ambient medium of the GRB. The flux density prediction in column (4) is calculated at $30~\text{MHz}$, for a burst like GRB 170112A with duration $0.06~\text{s}$ and fluence $0.13\times10^{-7}~\text{erg cm}^{-2}$, and assumes $\delta$, defined by the authors as the fluence in radio emission relative to the fluence measured in gamma-rays, as $\delta=10^{-4}$. See Equation 14 in~\citealt{Usov+Katz2000}.}
	\tablecomments{The estimates and parameters used to calculate the predicted flux density $S$ use the approximate values reported in the respective model papers as astrophysically plausible and/or most-likely values. We emphasize that many of the parameters estimated in these models are unknown and not constrained astrophysically, therefore the appropriate ranges for these values can vary by orders of magnitude. \\
				(1) Reference for model predicting coherent radio emission associated with GRB.\\
				(2) Onset of radio emission relative to onset of gamma-ray emission.\\
				(3) Duration of coherent radio emission, before dispersion.\\
				(4) Model-predicted flux density for a burst located at $D=1~\text{Gpc}$, the approximate distance upper limit to GRB 170112A.\\
				(5) Spectral index.\\
				(6) Magnetic field strengths required by model in the GRB progenitor in order to satisfy conditions necessary for producing coherent radio emission.\\
				(7) Luminosity ratio between energy released in radio frequencies to energy released in gamma-rays, during the prompt emission stage. See~\citealt{Palmer1993}.}
\end{deluxetable}

Searches for prompt, coherent radio counterparts to GRBs are made difficult by the need for observations that satisfy the requirements for high sensitivity at sufficiently low frequency, and are coincident with (or, dependent on the amount of dispersive delay, shortly after) the detection of the corresponding GRB. There have been many searches for prompt, coherent radio counterparts to GRBs to date, but none have yielded detections thus far~\citep[e.g.][see \citealt{Granot+vanderHorst2014} for brief summary]{Baird+1975, Inzani+1982, Koranyi+1995, Dessenne+1996, Benz+Paesold1998, Balsano1999, Bannister+2012}. Most recently, \citealt{Obenberger+2014} searched for prompt emission from 32 GRBs using the LWA1 Prototype All Sky Imager~\citep[PASI;][]{Ellingson+2013, Obenberger+2015}, and \citealt{Kaplan+2015} conducted follow-up observations of the short GRB~150424A, starting within 23~seconds of the detected gamma-rays, with the Murchison Widefield Array~\citep[MWA;][]{Tingay+2013} at frequencies above 80~MHz. See Table~\ref{tab:previousobservations} for a summary of previous radio surveys specifically targeting prompt emission associated with GRBs.

\begin{deluxetable}{ccCCCCCCC}[htb!]
	\tabletypesize{\scriptsize}
	\tablecolumns{9}
	\tablecaption{Searches targeting prompt, coherent radio emission associated with GRBs\label{tab:previousobservations}}
	\tablehead{ 
	(1)					& (2)					& (3)					& (4)					& (5)					& (6)					& (7)						& (8)						& (9)						\\
	\colhead{Reference(s)} 	& \colhead{Instrument} 	& \colhead{Frequency} 	& \colhead{Bandwidth}	& \colhead{Sensitivity} 	& \colhead{Time res.} 	& \colhead{Frequency res.} 	& \dcolhead{t_\text{on sky}} 	& \dcolhead{N_\text{GRBs}}	\\
					    	&  					& \colhead{[MHz]} 		& \colhead{[MHz]}		& \colhead{[Jy]} 		& \colhead{[s]} 			& \colhead{[MHz]} 			&  						& 						}

	\startdata
	\citealt{Baird+1975} 						& spaced receivers\tablenotemark{1}		& 151 		& 1		& 10^5 		& 0.3				& 0.2			& -1~\text{hr} 			& 19\\ 
	& \vspace{-0.3cm} VHF/UHF station &&&&&&& \\
	\vspace{-0.3cm} \citealt{Cortiglioni+1981}		& 								& 151,~408	& 1,~2	& 10^4		& 0.3				& \nodata		& -1~\text{hr}			& 32\\
	& at Medicina &&&&&&& \\
	& \vspace{-0.3cm} VHF/UHF station &&&&&&& \\
	\vspace{-0.3cm} \citealt{Inzani+1982}		& 					 			& 151,~408 	& 1	 	& 10^4 		& 1				& \nodata		& -1~\text{hr}	 		& 65\\ 
	& at Medicina &&&&&&& \\
	\citealt{Koranyi+1995} 					& CLFST		 					& 151 		& \nodata 	& 200		&1.5				& \nodata		& 1~\text{hr} 			& 1\\ 
	\citealt{Dessenne+1996} 					& CLFST		 					& 151		& 0.7		& 73,~35,~15 	&1.5				& \nodata		& -280,~6,~16~\text{m} 	& 2\\
	& \vspace{-0.3cm} solar radio &&&&&&& \\
	\vspace{-0.3cm} \citealt{Benz+Paesold1998} 	& 								& 40-1000		& \nodata	& 10^5		& 0.25			& 1			& \text{simultaneous}	& 7\\
	& spectrometers\tablenotemark{2} &&&&&&& \\
	\citealt{Balsano1999}					& FLIRT		 					& 74			& 1.9		& 10^3		& 0.05			& \nodata		& 10~\text{s}			& 32\\
	& \vspace{-0.3cm} $12$-m dish at &&&&&&& \\
	\vspace{-0.3cm} \citealt{Bannister+2012}		& 					 			& 1400		& 220	& 7			& 64\times10^{-6}	& 0.390		& 200~\text{s}			& 9\\
	& Parkes Observatory &&&&&&& \\
	\citealt{Obenberger+2014}				& LWA1							& 37.9,~52,~74	& 0.075	& 68,~65,~70	& 5				& 0.0167		& \text{simultaneous}	& 34\\
	\vspace{-0.3cm} & & 80,~88.9,~97.9, && 8.7,~7.7,~5.7, &&&& \\
	\vspace{-0.3cm} \citealt{Kaplan+2015}		& MWA						 	& 			& 2.56	& 			& 4				& \nodata		& 23~\text{s}			& 1\\
	& & 108.1,~119.7,~132.5 && 4.9,~4.2,~3.0 &&&& \\
	This Work								& OVRO-LWA					 	& 56			& 57		& 0.650		& 13				& 0.024		& -1~\text{hr}			& 1\\
	\enddata
	\tablenotetext{1}{The receivers were located in stations at Cambridge, UK; Dublin, Ireland; Glasgow, UK; Harwell, UK; Jodrell Bank, UK; and Malta.}
	\tablenotetext{2}{Located in Bleien Radio Observatory, Switzerland; Tremsdorf, Germany; and Weissenau Observatory, Germany.}
	\tablecomments{\\ (8) Start of observations, relative to the time of GRB. \\ (9) Number of GRBs targeted in the radio observations.}
\end{deluxetable}
 
Despite these difficult observational requirements, and the speculative nature of the models, low frequency radio searches for counterparts to short GRBs remain valuable. Systematic detection would provide a radio source population with the ability to probe the density and turbulence of the intergalactic medium~\citep[IGM;][]{Inoue2004}, with utility as a diagnostic of accretion-powered jet physics~\citep{Macquart2007}, and as a valuable electromagnetic (EM) counterpart for radio follow-up of gravitational wave (GW) events~\citep[e.g.,][]{Kaplan+2016}. Most critically, the detection of a radio pulse associated with a short GRB would provide independent confirmation of the association of short GRBs with neutron star mergers. This is especially relevant in the current era of multi-messenger astronomy ushered in by the detection of GW170817~\citep{Abbott+2017c}, which provided the long sought-after first direct association between gamma ray bursts and binary neutron star mergers. However, the gamma ray burst detected in association with GW170817 remains distinct from the ``classical" short GRBs that are systematically detected at larger distance~\citep{Abbott+2017d}. A coherent, low frequency radio counterpart to binary neutron star mergers would provide a direct link between short GRBs and their more local GW-detected counterparts.

We have conducted the most sensitive search to date at frequencies below 100 MHz for a prompt, coherent radio counterpart associated with GRBs, using the Owens Valley Radio Observatory Long Wavelength Array (OVRO-LWA) to observe the field of the short GRB 170112A. The low frequency ($27-85$~MHz) and simultaneous nature of our observations provides the most constraining limits on a number of the prompt radio emission counterpart models. In \S~\ref{observations} we describe the OVRO-LWA and our observations of GRB 170112A. In \S~\ref{analysis} we describe the analysis, including our dedispersion search. In \S~\ref{discussion} we place limits on any prompt radio emission associated with 170112A and the resulting constraints on the models, as well the relevance of the OVRO-LWA observations in the context of GW follow-up. We conclude in \S~\ref{conclusion}.

\section{Observations}\label{observations}

\subsection{OVRO-LWA}
The OVRO-LWA is a 352-element, dual-polarization dipole array currently under development at the Owens Valley Radio Observatory (OVRO) in Owens Valley, California~(Hallinan et al.~in prep), operating at frequencies below 100~MHz. The final array will be spread over a 2.5~km diameter area, providing a roughly 5~arcminute spatial resolution. Full cross-correlation of all 352 elements will enable imaging of the entire viewable sky, with a cadence of a few seconds and 100~mJy snapshot sensitivity. Early science observations have commenced on OVRO-LWA, with key science including low frequency radio transients, exo-space weather monitoring of nearby stellar systems, cosmic dawn 21~cm science~\citep[see][]{Price+2017, Eastwood+2017}, ionospheric studies, solar dynamic imaging spectroscopy, and monitoring of the Jovian system. A unique design feature of the array is the non-conflicting nature of these disparate science goals, which all share a common mode of observing and initial data products, meaning that all science objectives can be served simultaneously.

The current, stage II OVRO-LWA, which incorporates 256 elements from the 200~meter diameter core and the 32-element Long Baseline Demonstrator Array (LBDA), has been observing continuously as of December 2016, operating from 27-84 MHz (2400 channels) with a 13-second cadence. Full cross-correlation of 512 inputs (256 antennas $\times$ 2 polarizations) by the Large-Aperture Experiment to Detect the Dark Age (LEDA) correlator~\citep{Kocz+2015} provides a full-sky field of view with an approximately 10~arcminute resolution at the top of the observing band. Data are continuously written to a multi-day buffer. Data corresponding to triggers of interest are copied to the on-site All-Sky Transient Monitor (ASTM) for storage and processing. Example triggers of interest include \textit{Swift} and \textit{Fermi} alerts for both short and long GRBs as well as GW LIGO-Virgo Collaboration (LVC) events, as distributed by the Gamma-ray Coordinates Network\footnote{GCN: \url{https://gcn.gsfc.nasa.gov/}}. The short GRB 170112A represents the first short GRB with sufficient position localization to search for a coherent radio emission counterpart with the OVRO-LWA, following the onset of our stage II continuous mode of observing. The results of follow-up observations of a larger sample of both long and short GRBs with known redshift will be released once the stage II OVRO-LWA continuous operations have completed.

\subsection{Short GRB 170112A}
The short GRB 170112A was detected on 2017 January 12 02:02:00 UTC by the \text{Swift} Burst Alert Telescope~\citep[BAT;][]{Gehrels+2004, Krimm+2013} at the position (RA, dec)~=~(01$^\text{h}$00$^\text{m}$55$^\text{s}$.7, -17$^\circ$13$'$57.9$''$), to within a $90\%$ error region of radius 2.5~arcminutes \citep{Mingo+2017, Lien+2017}. The burst was identified as a hard, short burst (power-law index $\alpha_{\text{PL}}=-1.2$ and burst fluence $S=0.13\times10^{-7}$~erg~cm$^{-2}$ in the \textit{Swift}-BAT 15-150~keV band), with an atypically short duration of $T_{90}=0.06$~seconds~\citep[][see discussion in \S~\ref{subsec:dispersionandscattering}]{Lien+2017}. No extended emission was found, as is typical of the majority of short GRBs~\citep{Bostanci+2013}. Due to the lack of detected X-ray flare or afterglow emission, by either the \textit{Swift} X-Ray Telescope~\citep[XRT;][]{Dai+2017} or \textit{Swift} Ultraviolet/Optical Telescope~\citep[UVOT;][]{Siegel+Mingo2017}, following the initial BAT trigger, the position of 170112A is known only to within the 2.5~arcminute position as measured by BAT. No afterglow counterpart or associated host galaxy was found in the observations conducted by follow-up ground-based optical and NIR facilities. \citealt{DAvanzo+2017} observed the location of GRB 170112A with the REM~60-cm robotic telescope at La Silla Observatory in Chile, and placed an 18.0 magnitude H band upper limit on an afterglow counterpart. \citealt{Mazaeva+2017} report the detection by the 0.7-m AS-32 telescope at Abastumani Observatory of a source, inside the BAT error region, which is not present in the USNO-B1.0 catalog but is detected in the DSS2 (red) survey. We therefore disregard the association of this source with GRB 170112A, and in the remaining analysis consider GRB 170112A as a short burst with neither detected afterglow emission nor associated host galaxy.

The OVRO-LWA was observing simultaneous to the detection of 170112A by \textit{Swift} BAT. Following the GCN detection notice, and after receiving the verification by the GCN circular~\citep{Mingo+2017} that \textit{Swift} had detected a short GRB, we saved three hours of data from the OVRO-LWA transient buffer, corresponding to one hour prior to and two hours after the \textit{Swift} detection (UT range 01:02:02 -- 04:02:18). The data consist of 832 contiguous integrations of 13-second duration, with 2398 frequency channels spanning 27.38 -- 84.92 MHz (24 kHz frequency resolution). The OVRO-LWA is a zenith-pointing telescope, and 170112A was located at an elevation of approximately $35^\circ$ in the primary beam.

The visibility data delivered from the LEDA correlator are converted from their raw format into the standard CASA visibility table measurement sets. The data are then flagged in frequency, antenna, and baseline space, using a combination of manual inspection and an automated in-house algorithm that fits for smoothness in the visibilities across frequency, time, and \textit{uvw}-space~(Monroe et al.~in prep). Approximately 45\% of visibilities are flagged per integration. The data are calibrated using a simplified sky model consisting of the two brightest sources in the sub-100~MHz sky: the radio galaxy Cygnus (Cyg) A and the supernova remnant Cassiopeia (Cas) A. The complex (amplitude and phase) antenna gains are determined on a per-channel basis from the Cyg A-Cas A sky model using the CASA \texttt{bandpass} task. The calibration solutions are derived from an integration taken at 2017-01-11 20:26 UT, 4 hours prior to the start of the GRB follow-up observations, when Cyg A is at its highest elevation in the beam ($\sim87^\circ$). The flux scale and bandpass are corrected by modifying the antenna gain amplitudes to fix the spectrum of Cyg A to that of \citealt{Baars+1977}. Direction-dependent calibration in the directions of Cyg A and Cas A is performed using the \texttt{TTCal} calibration software package developed for the OVRO-LWA~\citep{eastwood_michael_w_2016_1049160}. This is necessary in order to accurately peel the sources from the visibilities and avoid sidelobe contamination in the images, caused by antenna gain pattern variations between individual dipoles. Imaging and deconvolution are performed with \texttt{WSClean}~\citep{Offringa+2014}. The full field of view is imaged over 4096 $\times$ 4096 pixels, with a pixel scale of 1.875$'$ and using a robust visibility weighting of 0~\citep{Briggs1995}. At the time of these observations, a set of northeast LBDA antennas were not operational, resulting in an abnormally elongated synthesized beam with a major axis of $29'$, a minor axis of $13.5'$, and a position angle of $50^\circ$.

\begin{figure*}[htb!]
\begin{center}
	\subfigure[]{\includegraphics[width=0.5\textwidth]{f1a.pdf}}
	\subfigure[]{\includegraphics[width=0.49\textwidth]{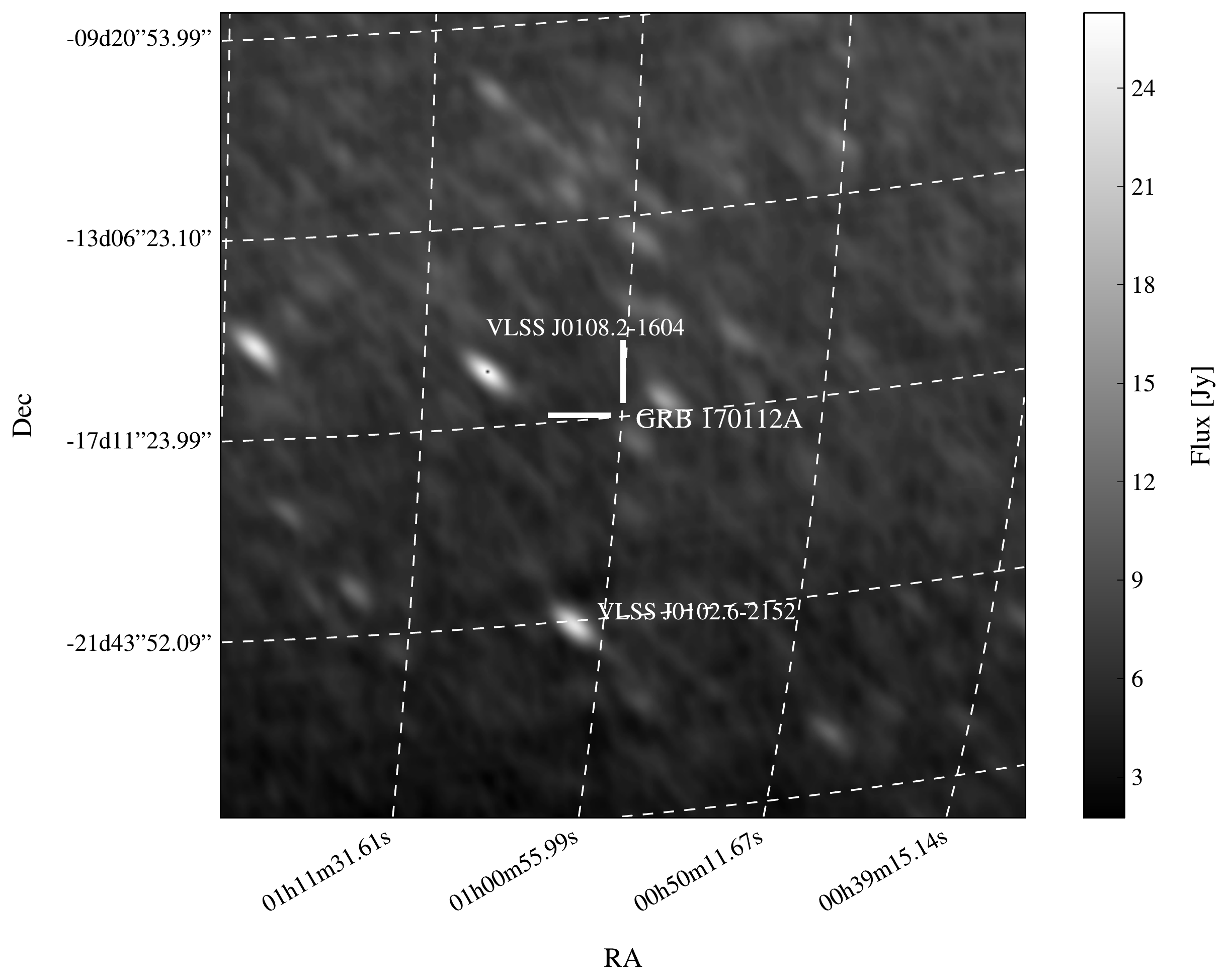}}
	\caption{The left panel is a 13-second, fullband snapshot OVRO-LWA image corresponding to the time of the GRB detection by \textit{Swift}, at 02:02:05 UT. The center of the image is zenith and the border of the image is the horizon-line. The extended emission cutting across the top half of the image is predominantly synchrotron emission from our own galaxy. Cas A and Cyg A have been peeled from this integration. There are roughly 10,000 point sources in this 13-second image. The noise at zenith is approximately 800~mJy. The square box at $35^\circ$ elevation in the southwest of the image corresponds to the right panel, which shows a $10^\circ\times10^\circ$ box, centered on the location of 170112A at (RA, dec)~=~(01$^\text{h}$00$^\text{m}$55$^\text{s}$.7, -17$^\circ$13$'$57.9$''$). VLSS J0108.2-1604 and VLSS J0102.6-2152 are also labeled.}
	\label{fig:fullsky}
\end{center}
\end{figure*}

Figure \ref{fig:fullsky} shows the 13-second, fullband snapshot OVRO-LWA image corresponding to the time of the GRB detection by \textit{Swift}. The images are confusion-noise limited to $\sim 800$~mJy at zenith. The image cutout in Figure \ref{fig:fullsky} shows the $10^\circ\times10^\circ$ region surrounding the position of 170112A, including two nearby VLA Low Frequency Sky Survey~\citep[VLSS;][]{Lane+2014} sources, VLSS  J0108.2-1604 (3C 032), and VLSS J0102.6-2152. These were used to verify the flux scale at this part of the primary beam and track the position offset of sources over the course of the observation caused by ionospheric refraction. An ionospheric substorm can be seen moving through this region of the sky, particularly during the first hour of observations, however the maximum source position offsets are within the size of the synthesized beam, and on average are of order a few arcminutes.

\section{Analysis}\label{analysis}
\begin{figure*}[htb!]
\begin{center}
	\includegraphics[width=0.8\textwidth]{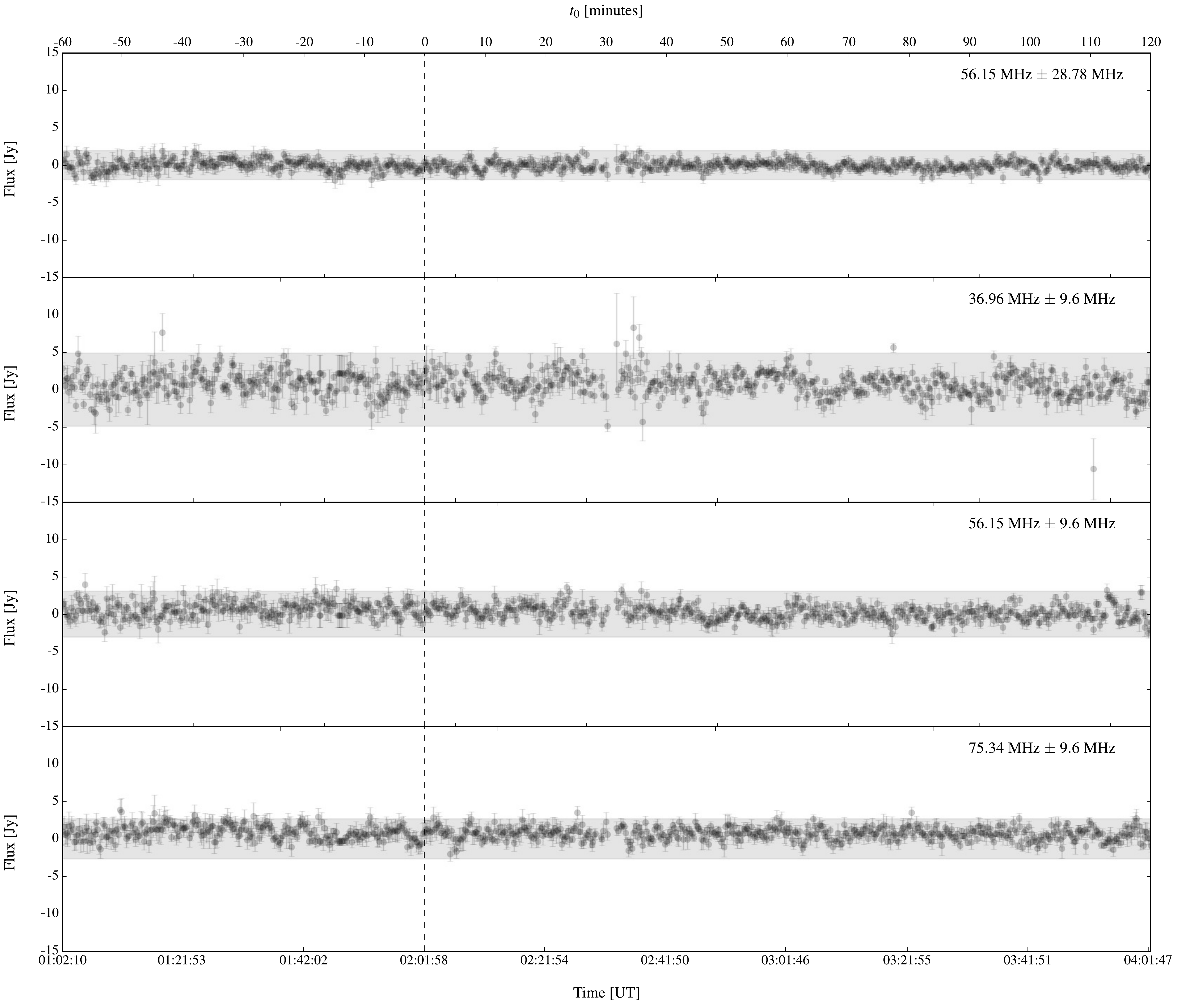}
	\caption{Flux density at the position of short GRB 170112A in each 13-second integration, at 56~MHz with the full 57~MHz bandwidth (top), the bottom third of the band centered at 37~MHz (second from top), the middle third of the band at 56~MHz (third from top), and the top third of the band at 75~MHz (bottom). The time series shows the full three-hour observation, starting one hour prior to the \textit{Swift} detection of 170112A at $t_0=0$ (dashed line) and ending two hours later. The noise in each band is represented by the gray regions ($3\sigma$), with noise in each band of 650~mJy, 1.6~Jy, 1.0~Jy, and 890~mJy, respectively. We detect no statistically significant emission indicative of prompt radio emission associated with the short GRB.}
	\label{fig:timeseries}
\end{center}
\end{figure*}

To search for the presence of a low frequency counterpart to 170112A in our data, we searched the flux density time series at the position of 170112A, known to within one synthesized beam, for statistically significant peaks indicative of a radio burst. In each integration, the median flux of an annulus around the GRB position of width $6\pm1$ synthesized beams was subtracted from the flux measured at the pixel corresponding to the position of GRB 170112A in order to remove additional flux from any large-scale, diffuse structure. This was done for the full 57~MHz bandwidth, as well as the bottom (centered at 37~MHz), middle (56~MHz), and top (75~MHz) thirds of the band. Figure \ref{fig:timeseries} shows the time series for these four bands for the duration of the three hour observation. We find no statistically significant peaks in any of the four bands, with typical noise in the full band, bottom band, middle band, and top band light curves of 650~mJy, 1.6~Jy, 1.0~Jy, and 890~mJy, respectively.

\subsection{Pulse Propagation: Dispersion and Scattering}\label{subsec:dispersionandscattering}
The propagation of a coherent radio pulse associated with 170112A would be affected by the cold plasma in the intervening medium between the birth site of the pulse and the Earth. The signal will be dispersed, with lower frequencies arriving at later times relative to higher frequencies, following $t_\text{arrival} = 4.2~\text{DM}~\nu_\text{GHz}^{-2}~\text{ms}$~\citep{Cordes+McLaughlin2003}. The amount of dispersion, quantified by the dispersion measure (DM), is determined by the column density of electrons along the pulse path (density of electrons integrated along the line of sight), and is equal to $\text{DM} = \int_0^D n_e(l)dl$, where $D$ is the distance to the source and $n_e$ is the density of electrons. The DM can be approximated as the sum of dispersion measure contributions from the Milky Way ($\text{DM}_\text{MW}$), the intergalactic medium ($\text{DM}_\text{IGM}$), and the host galaxy of the burst ($\text{DM}_\text{host}$). The DM contribution from the Milky Way is minimal due to the location of 170112A at high Galactic latitude ($l=135.9^\circ$, $b=-79.9^\circ$); $\text{DM}_\text{MW} \sim 30$~pc~cm$^{-3}$ based on the NE2001 model of \citealt{Cordes+Lazio2002}\footnote{\url{https://www.nrl.navy.mil/rsd/RORF/ne2001/}}. We assume that the contributions from the host galaxy and circumburst environment of 170112A are similarly small, given the likely binary NS-NS(BH) progenitors of short GRBs and the typically large ($4.5~\text{kpc}$ projected median) offsets from the center of their hosts galaxies~\citep{Fong+Berger2013}. We assume a $\text{DM}_\text{host} \approx \text{DM}_\text{MW}$. The DM contribution from the IGM can be estimated as $\text{DM}_\text{IGM} \sim 1000~z$~pc~cm$^{-3}$~\citep{Zheng+2014}. Because 170112A has no optical afterglow detection and no identified host galaxy, and thus is of unknown redshift, $\text{DM}_\text{IGM}$ is highly uncertain. However, we exploit the relationship between redshift and minimum burst duration of the prompt gamma-ray emission ($T_{90}$) determined using the sample of \textit{Swift}-detected GRBs with redshift identifications from the third \textit{Swift} BAT GRB catalog~\citep{Lien+2016} to place an upper limit on the redshift of 170112A of $z\lesssim 0.2$. The correlation between minimum detectable $T_{90}$ and redshift found by~\citealt{Lien+2016} reflects that longer exposure times, and therefore longer burst durations, are needed to detect lower flux bursts. The three-burst sample with $T_{90} < 0.1$~s all having redshifts $z \lesssim 0.2$ indicates that extremely short bursts must be brighter (and thus very nearby) to be detected. However, this correlation is necessarily restricted to bursts of known redshift, and we state the caveat that bright, short bursts at high redshift may indeed be detected but are simply lacking in redshift measurements. Based on the assumptions and estimates given above, we assume an upper limit of $\text{DM} \lesssim 260~\text{pc cm}^{-3}$. This corresponds to a maximum dispersive delay of $\sim 1330~\text{s}$ across the full OVRO-LWA $57~\text{MHz}$ bandwidth, and a maximum arrival time at the top of the band of $155~\text{s}$ after the GRB, indicating our observation covering $2$ hours post-GRB-detection is more than sufficient for capturing a coincident radio burst. The corresponding dispersion smearing across the $24~\text{kHz}$ channel at the bottom of the frequency band is $\tau_\text{DM} = 8.3 \times 10^3~\text{s} \left( \Delta\nu/\text{MHz} \right) \left( \nu_c/\text{MHz} \right)^{-3} \left( \text{DM}/\text{pc}~\text{cm}^{-3} \right) \approx 2.5~\text{s}$, shorter than the integration time $t_\text{int}=13~\text{s}$ of these observations.

\begin{figure*}[htb!]
\begin{center}
	\includegraphics[width=1\textwidth]{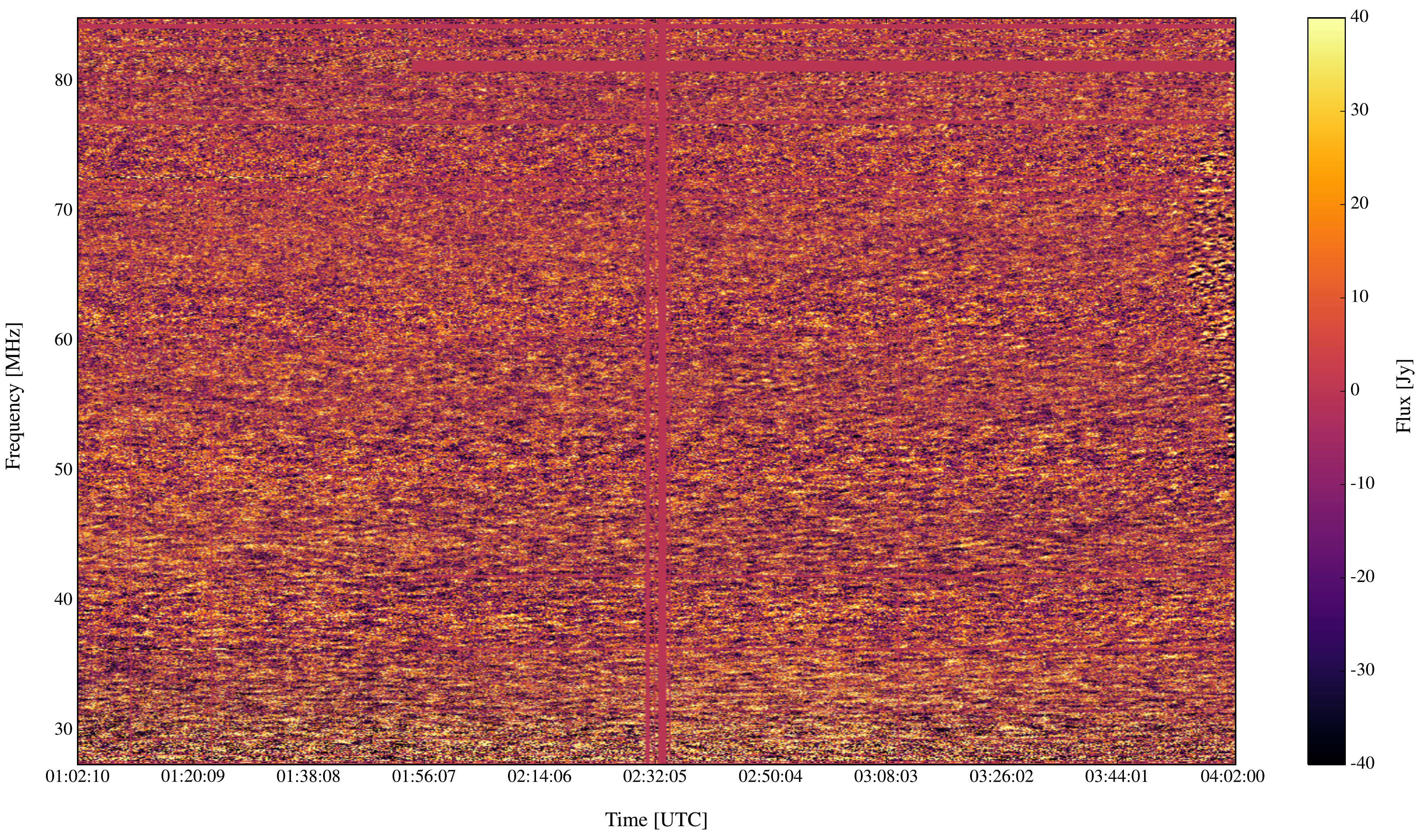}
	\caption{The dynamic spectrum measured at the position of GRB 170112A. The data span $1$ hour prior to and $2$ hours following the gamma-ray emission, at a time resolution of 13 seconds. The frequency channel width is $24~\text{kHz}$, with $2398$ channels spanning $27.4$ through $84.9~\text{MHz}$. Some frequency and time bins have been flagged due to RFI (e.g., integrations surrounding 02:32:05 UTC).}
	\label{fig:dynspec}
\end{center}
\end{figure*}

Figure \ref{fig:dynspec} shows the undispersed dynamic spectrum at the location of GRB 170112A, formed by measuring the flux at the location of GRB 170112A in the $24~\text{kHz}$-wide channel images across the full bandwidth, for every $13~\text{second}$ integration across the duration of the observation ($832$ integrations in total). Similar to the process by which the time series of Figure \ref{fig:timeseries} were formed, we subtracted the median flux of an annulus around the GRB position of width $6\pm1$ synthesized beams from the flux measured at the pixel corresponding to the position of GRB 170112A in order to remove additional flux from any large-scale, diffuse structure. We performed a series of dedispersion trials on the dynamic spectrum, with the spacing between DM trials set by the amount of dispersion smearing acceptable across the full bandwidth in the final, dedispersed time series. We set this equal to the integration time $t_\text{int} = 13~\text{s}$. Following from
\begin{equation}
\Delta t_\text{smearing} = t_\text{int} = 4.15~\text{ms}~\text{DM}_\text{step} \left[ \frac{1}{\nu_\text{1,GHz}^2} - \frac{1}{\nu_\text{2,GHz}^2} \right],
\end{equation}
this gives a DM step size of $\text{DM}_\text{step} = 2.5~\text{pc cm}^{-3}$. We search DMs spanning $0$ through $1000~\text{pc cm}^{-3}$, despite our assumption of low DM, so as not to preclude the possibility of larger distances to the source or a significant contribution from the host galaxy or circumburst medium.

\begin{figure*}[htb!]
\begin{center}
	\includegraphics[width=1\textwidth]{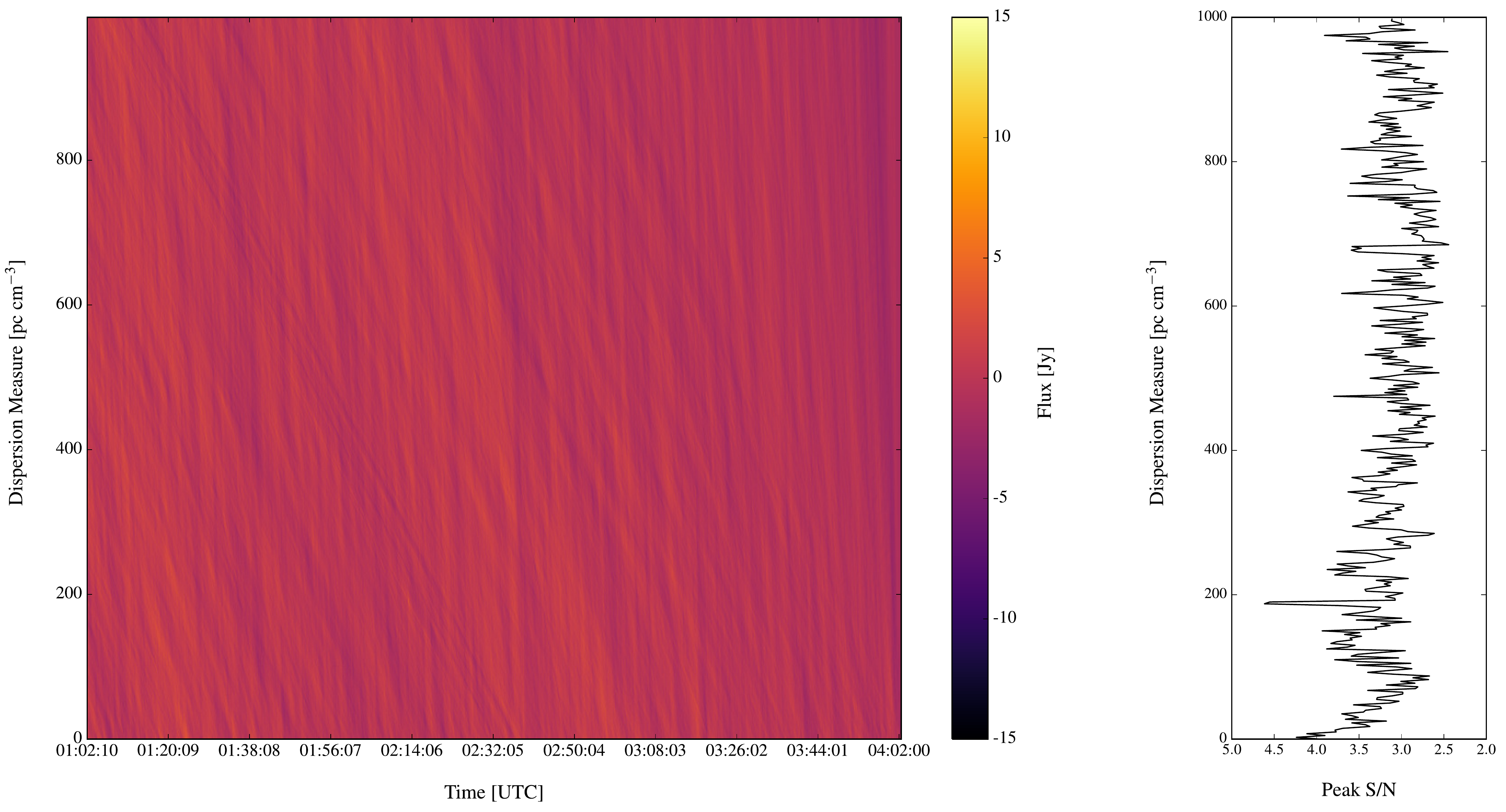}
	\caption{Dedispersed time series for every DM trial ranging from $0$ to $1000~\text{pc cm}^{-3}$, at $2.5~\text{pc cm}^{-3}$ intervals. The right panel shows the peak S/N in each dedispersed time series as a function of DM trial. No time series contains a peak S/N greater than our significance threshold. The $4.5\sigma$ spike just below a DM of $200~\text{pc cm}^{-3}$ is $2$ DM channels wide, and is due to spurious unflagged frequency channels.}
	\label{fig:dmchan}
\end{center}
\end{figure*}

In order to search for the presence of a pulse in each of the trial DM dedispersed time series, we searched for statistically significant emission in every time series trial.
Figure \ref{fig:dmchan} shows the full-band dedispersed time series for every DM trial. We assume that our coarse $13~\text{second}$ integration time resolution is greater than the intrinsic pulse width (see Table \ref{tab:models}), as well as any additional broadening due to scattering or remaining dispersion smearing due to the finite spacing of our DM trials, thus no additional smoothing is performed on the time series. Our non-detection of a radio pulse is consistent with the flat behavior of S/N we see as a function of DM channel.

In addition to the removable effect of dispersive delay caused by the intervening plasma along the path of pulse propagation, the signal is also subject to non-removable scatter broadening by turbulence and plasma density variations along the line of sight. The contribution to scattering from the Milky Way is negligible~\citep[a pulse broadening timescale of $\tau=0.06~\mu\text{s}$ at $1~\text{GHz}$ will be $\tau < t_\text{int} = 13~\text{s}$ at $10~\text{MHz}$, assuming a Kolmogorov $\nu^{-4}$ power-law;][]{Cordes+Lazio2002}. Additional contributions to scattering are possible from the host galaxy, however for the same reasons mentioned above with regards to host galaxy contribution to DM, we assume a negligible contribution from the host, given the likely significant offset of the short GRB from its host galaxy. The remaining source of scattering to be considered is the IGM. Given both the lack of correlation between DM and $\tau$ for the extragalactic fast radio bursts (FRBs) and their systematically under-scattered, relative to the Galactic DM-$\tau$ relation for pulsars, nature~\citep{Cordes+2016}, we assume a minimal contribution to scattering from the IGM (which we have taken as the largest contribution to the DM), and assume $\tau < t_\text{int}$.

\section{Discussion}\label{discussion}
\subsection{Constraints on Models for Prompt Radio Emission}
\begin{figure*}[htb!]
\begin{center}
	\subfigure[]{\includegraphics[width=0.65\textwidth]{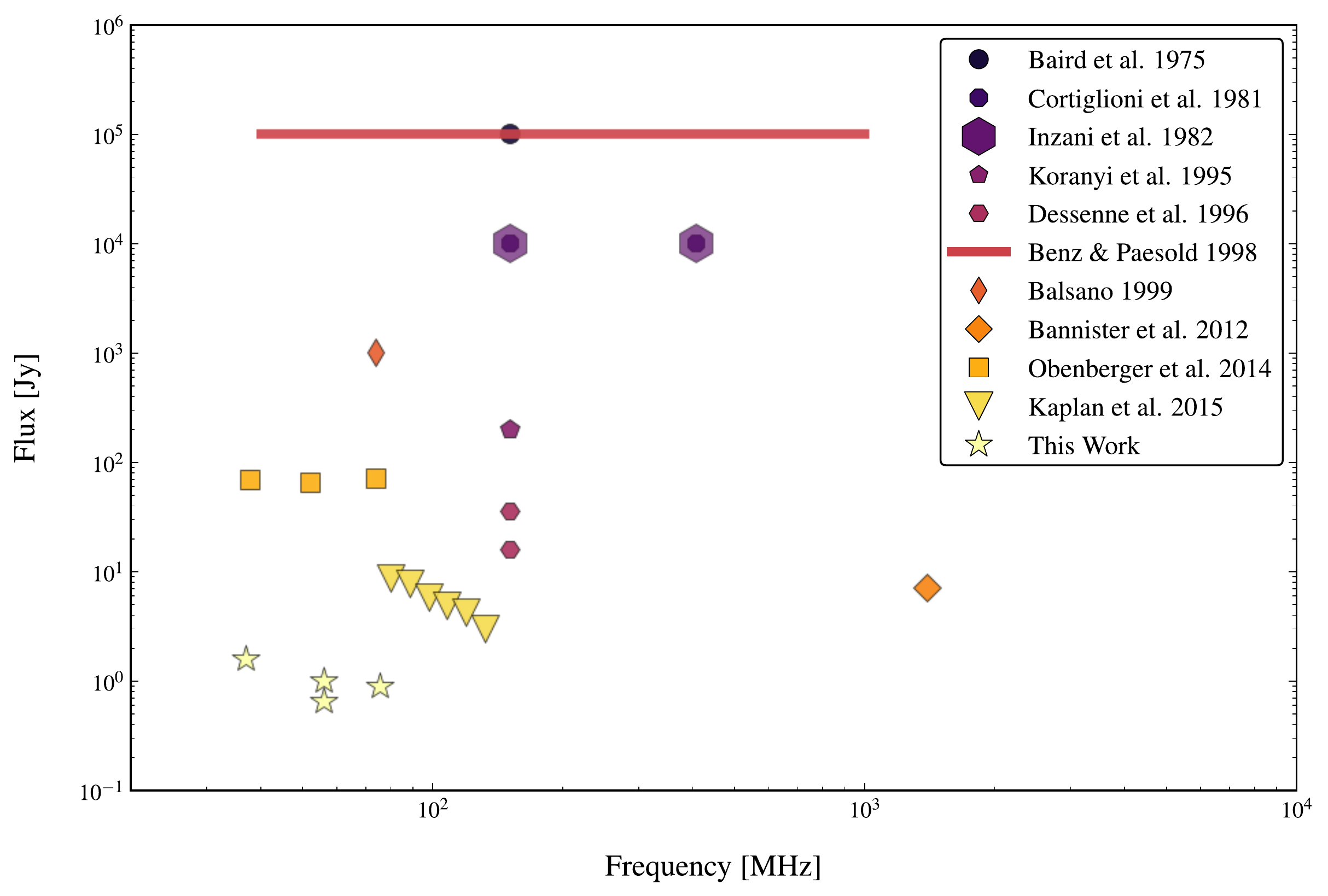}}
	\subfigure[]{\includegraphics[width=0.65\textwidth]{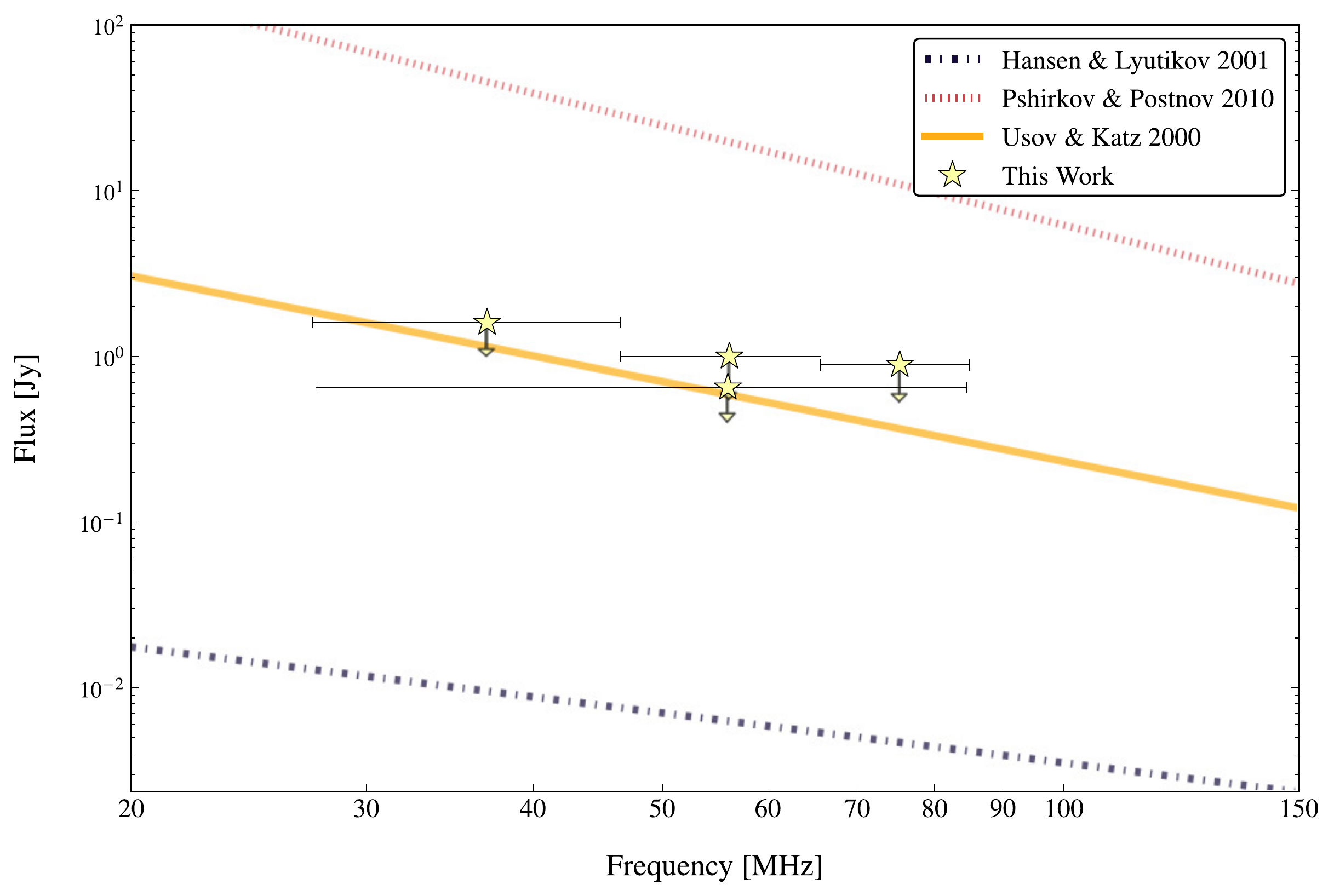}}
	\caption{Top: Flux density limits from all previous searches targeting prompt coherent radio emission associated with GRBs. See also Table \ref{tab:previousobservations}. Surveys reporting limits at multiple frequencies show the corresponding number of limits in the plot. Bottom: The $1\sigma$ flux density limits from the full $57~\text{MHz}$ band and the three $19~\text{MHz}$ subbands for GRB 170112A, and the model predicted flux densities from Table \ref{tab:models}, scaled to the $13~\text{s}$ integration time of these observations.}
	\label{fig:limits}
\end{center}
\end{figure*}

Using our most constraining limit on prompt radio emission associated with the short GRB 170112A ($57~\text{MHz}$ bandwidth flux density limit at a center frequency of $56~\text{MHz}$), we can place limits on the efficiency factors (listed in column 7 of Table \ref{tab:models}) used in the models for prompt radio emission to predict the strength of emission, as well as place an upper limit on the fractional energy released in low frequency radio emission relative to gamma-rays in the prompt emission stage of a short GRB. Figure \ref{fig:limits} shows the predictions for the prompt radio emission models of \citealt{Hansen+Lyutikov2001}, \citealt{Pshirkov+Postnov2010}, and \citealt{Usov+Katz2000}, associated with a 170112A-like burst, compared with the flux density $S(56~\text{MHz}) = 650~\text{mJy}$ of our most constraining limit (assuming a burst at a distance of $D=1~\text{Gpc}$, and the model-predicted spectral indices $\alpha$ given in column 5 of Table \ref{tab:models}). For the magnetized wind resulting from the magnetospheric interaction of the binary neutron stars prior to coalescence model of \citealt{Hansen+Lyutikov2001}, we place an upper limit on the efficiency of wind power conversion to radio emission of $\epsilon_R \lesssim 10^{-3}$ (assuming a magnetic field strength of $B=10^{12}~\text{G}$ for the non-magnetar component of the BNS system considered in the model). For the pulsar-like coherent radio emission powered by the rapid spin-down of the post-merger supramassive magnetar model of \citealt{Pshirkov+Postnov2010}, we place an upper limit on the efficiency with which spin-down energy is converted into coherent radio emission of $\eta \lesssim10^{-8}$, corresponding to a total energy loss rate of roughly $\dot{E}=3\times10^{51}~\text{ergs s}^{-1}$ and a power-law for the spin-down power ($\eta \propto \dot{E}^\gamma$) of $\gamma = -0.2$. For the magnetized outflow generating coherent radio emission simultaneous to the gamma-ray emission powering the GRB model of \citealt{Usov+Katz2000}, we place an upper limit on the ratio of emitted radio to gamma-ray fluence of $\delta \lesssim 10^{-4}$ (using the measured fluence of GRB 170112A in the $15-150~\text{keV}$ band of $0.13\times10^{-7}~\text{erg cm}^{-2}$). If we assume a similar time-scale for the emission of both prompt radio and gamma-ray emission, and a spectral index $\alpha = -2$, using the gamma-ray fluence of GRB 170112A, we can place an upper limit on the fraction of energy released in the prompt radio emission (across our $27$ to $85~\text{MHz}$ band) relative to gamma-rays of $L_\text{radio}/L_\gamma \lesssim 7\times10^{-16}$.

\subsection{Gravitational Wave Follow-up}
Verifying the existence of a prompt low frequency radio counterpart to NS-NS(BH) merger events has taken on new significance with the detection of the binary neutron star merger GW170817 in gravitational waves and the detection of its counterparts across the electromagnetic spectrum~\citep{Abbott+2017e, Abbott+2017c}. Strategies for follow-up of GW events with EM facilities are now being regularly deployed as GW alerts are released to partner observers and facilities, spanning the entire EM spectrum from radio to gamma-rays, and including neutrino facilities~\citep{Singer+2014}. Though the current models predicting prompt radio counterparts to both short and long GRBs remain very speculative, the utility of such a population for constraining important jet physics~\citep{Macquart2007}, probing the IGM~\citep{Inoue2004}, and serving as an identifiable counterpart to GW events makes the search for such events extremely valuable~\citep[e.g.,][]{Kaplan+2016}. The particular value of a low frequency prompt counterpart to NS-NS(BH) mergers over counterparts at higher energies is the (1) inherently wide-field of view of low frequency facilities for rapidly and efficiently covering the entire (often $>10^3~\text{sq.~deg}$) localization regions of GW events detected by aLIGO; (2) the time delay of as much as minutes, caused by the dispersion of a radio pulse by the media along the path of propagation, that allows for the pointing of telescopes; and (3) the relative dearth of variable and un-related transient sources which could serve as false positives~\citep[unlike at optical frequencies; see e.g.][and references therein]{Bhalerao+2017}.

The OVRO-LWA in particular is a uniquely powerful follow-up facility for GW transients. The nearly full-hemisphere ($\sim$20,000 sq.~deg.) field of view can instantaneously cover the entire $90\%$ confidence localization region released by the two aLIGO and Virgo detectors roughly 50\% of the time. The dispersive delay of a prompt radio counterpart enables timely follow-up by a number of low frequency facilities. However, the benefits of dispersive delay become less relevant as latency between GW detection and notification of detection to partner EM facilities increases. During the first aLIGO observing run (O1), the median latency between detection and notification was approximately $60~\text{seconds}$, however the largest latencies extended beyond $180~\text{seconds}$~\citep{Abbott+2016}. For GW170104, detected during the O2 run, the latency was approximately $5~\text{hours}$~\citep{Abbott+2017,LIGO+2017}. However, large latency is not a limiting factor in the follow-up strategy of the OVRO-LWA, where the continuous mode of operation allows for the recovery of data from within the ring buffer, enabling contemporaneous coverage of a GW event and making possible the detection of any precursor radio emission, provided only that the GW alert latency is less than the duration of the OVRO-LWA buffer (which is typically $>24~\text{hours}$).

The OVRO-LWA was observing continuously during the aLIGO O2 observing run, but was unable to search for prompt coherent emission associated with GW170817 due to the unlucky placement of the binary neutron star merger below the horizon of the OVRO-LWA at the time of the event. GW170817 was located at a distance of only $40~\text{Mpc}$~\citep{Abbott+2017e}, and coincident observations of a burst at this distance would have placed the model flux density predictions in Figure~\ref{fig:limits} well above the current OVRO-LWA flux density limits and allowing significant constraints on the validity of the models. The detection horizon for the Advanced LIGO and Virgo detectors during the O3 run to NS-NS(BH) mergers will be $220(400)~\text{Mpc}$~\citep[averaged over direction and inclination angle of system,][]{Abadie+2010}, still well within the $1~\text{Gpc}$ distance upper limit to GRB 170112A, and still closer than any short GRB of known redshift detected through prompt gamma- or X-ray emission. Any burst occurring within that horizon and detected by \textit{Swift} or \textit{Fermi} with unknown redshift would had to have been extremely under-luminous compared to any burst of known redshift; indeed, this was the case with GRB 170817A, which was identified in association with GW170817 but was extremely under-luminous and below the standard detection threshold of \textit{Fermi}~\citep{Abbott+2017d}. The implication of a population of nearby merging BNS systems that do not follow the same brightness distribution or emission characteristics in their prompt high energy emission compared to their higher redshift counterparts underscores the value of an EM signature and emission mechanism that is detectable at low frequencies, for characterizing and providing an additional EM counterpart for a GW NS-NS(BH) coalescence event. The OVRO-LWA will continue the program for monitoring and triggering on GW events during the O3 run which is expected to detect between $\sim6-120$ binary neutron star merger events per year at design sensitivity.

\section{Conclusion}\label{conclusion}
We have placed the most constraining upper limits to-date on prompt radio emission associated with GRBs using the OVRO-LWA to observe the field of short GRB 170112A. We searched for a pulse at the location of GRB 170112A starting $1~\text{hour}$ prior to the detection of gamma-ray emission by \textit{Swift} through the subsequent $2~\text{hours}$, to accommodate even extremely conservative estimates for pre-merger coherent radio emission models as well as the expected dispersive delayed onset of radio emission. We see no evidence for emission in the full-band ($\Delta\nu = 57~\text{MHz}$) time series or following a search through the dedispersed time series for DMs ranging from $0-1000~\text{pc cm}^{-3}$. We place a $1\sigma$ flux density limit of $650~\text{mJy}$ at $56~\text{MHz}$, which, assuming a nominal distance of $D=1~\text{Gpc}$ to 170112A, allows us to place upper limits on the efficiency factors required by the models to predict the luminosity of coherent radio emission associated with GRBs. We place an upper limit on the fraction of energy released in the prompt radio emission (across our $27$ to $85~\text{MHz}$ band) relative to gamma-rays of $L_\text{radio}/L_\gamma \lesssim 7\times10^{-16}$.

The contemporaneous follow-up of and sensitivity limits placed for the short GRB 170112A demonstrate the capabilities of the OVRO-LWA for targeted follow-up, particularly with regards to GW events. The OVRO-LWA is the only facility with the ability to instantaneously cover up to the full aLIGO / Virgo localization region with zero latency, and with typical sensitivities of $\sim 800~\text{mJy}$ at low frequencies. In addition, a GRB follow-up campaign with the OVRO-LWA is ongoing, and will provide limits on prompt radio emission for a sample that includes both long and short GRBs, in order to encompass a wide range of potential progenitor parameters.

\acknowledgements This material is based in part upon work supported by the National Science Foundation under Grant AST-1654815 and AST-1212226. GH acknowledges the support of the Alfred P.~Sloan Foundation and the Research Corporation for Science Advancement. The OVRO-LWA project was initiated through the kind donation of Deborah Castleman and Harold Rosen.

Part of this research was carried out at the Jet Propulsion Laboratory, California Institute of Technology, under a contract with the National Aeronautics and Space Administration, including partial funding through the President's and Director's Fund Program.

\bibliography{references}

\end{document}